\documentstyle[sprocl_josie]{article}
\def\ra{\rightarrow}
\def\be{\begin{equation}} 
\def\ee{\end{equation}}
\def\dM{\delta M}
\def\sl{\stackrel{\sim}{<}}
\def\s{\sim}
\begin{document}
\
\begin{flushright} 
UH-511-871-97   \\  
May 1997   
\end{flushright}
\title{FLAVOR CHANGING NEUTRAL CURRENTS IN CHARM SECTOR\footnote{Invited talk presented
at FCNC 97, Santa Monica, CA, Feb. 19-21, 1997, to be published in the Proceedings.} \\
       (as signal for new physics)}

\author{ S. PAKVASA}
\address{Department of Physics \& Astronomy \\
 University of Hawaii at Manoa \\
              Honolulu, HI 96822}
\maketitle


\section{Introduction}

I would like to discuss $D^0-\bar{D}^0$
the mixing and rare D decays as manifestations of Flavor Changing
Neutral Currents (FCNC) in the charm sector.  I will first review the expectations in the 
Standard Model (SM) and then summarize some typical expectations in new physics scenarios
\cite{Bigi1}.
I would like to argue that the charm case offers a large window of opportunity and it may be possible to learn something about the origin of the fermion mass matrix.

\section{$D^0 - \bar{D}^0$ Mixing} 

$D^0 - \bar{D}^0$ mixing differs from $K^0-\bar{K}^0$ and $B^0-\bar{B}^0$ mixing in 
several ways \cite{Burdman1}.  In the box diagram, the s-quark 
intermediate state dominates; this is in spite of 
the suppression by the factor 
$(m_s/m_c)^2$ resulting from the external momenta (i.e. the 
fact that $m_c > m_s)$\cite{Datta}.  The final result for $\delta m$ from 
the box diagram is extremely small, one finds

\be
\delta m_D \sim \quad 0.5.10^{-17} \ \ GeV
\ee
for  $m_s \sim 0.2$ GeV and $f_D \sqrt{B_D} \sim 0.2$ GeV; 
leading to

\be
\delta m_D/ \Gamma{_{D^{^0}}} \ \sim \ 3.10^{-5}
\ee
Although (or rather because) the short distance box diagram gives such a small
value, it has been a long-standing concern that the long distance effects may enhance 
$\dM$ considerably \cite{Donoghue}.  
We have begun a systematic dispersive approach; evaluating contributions
from single particle intermediate states, two particle intermediate states and so
on \cite{Burdman1}.  
We have found that 
$(\dM)_{1p} \sim 0.4.10^{-16}$ GeV and $(\dM)_{2p}$ (due to $p^+ p^-$ states) 
$\sim10^{-16}$ GeV.  It can be shown 
that the contributions from PV, VV and multiparticle states are
kinematically suppressed further.  In absence of conspiracies, we conclude 
$\dM \s 10^{-16}$ GeV.
Georgi and collaborators apply HQET \cite{Georgi} 
to the matrix element:  assume that $m_c$ is much larger
than typical hadronic scale, match the effective low energy theory at $m_c$ and  then run to low
energies.  No new operators arise and all long distance effects should come from the running.
The only operators then are the 4 quark operator yielding the usual box result; a 
6-quark operator which is about 3 times the box and an 8-quark operator which 
is about half the box.  The net result is a moderate $(\sim 3-4)$ enhancement with
$\dM \s 10^{-16} GeV$ in agreement with the dispersive estimate above.  Hence the SM expectation
for $\dM$ including long distance effects, is
\be
\dM \s 10^{-16} \ \mbox{GeV}
\ee
and hence  $x = \dM/\Gamma \sl 10^{-4}$.  We expect $\delta \Gamma$ to be of the 
same order as $\dM$ and hence $y = \delta \Gamma/2\Gamma \sl 10^{-4}$.  The SM expectation
for the mixing parameter $r_{mix}$ given by
\be
r_{mix} = \frac{x^2 + y^2}{2 + x^2 + y^2}
\ee
is $(r_{{mix})_{SM}} \sl  10^{-8}$.  Hence there are more than 5 orders of magnitude to search
for new physics (the current bound \cite{Tripathi} on $r_{mix}$  is $5.10^{-3})$.

CP violation in mixing can be described 
by two parameters related to the conventional $p$ and $q$:
\begin{eqnarray}
\frac{2Re \ \epsilon_D}{1 + \mid \epsilon_D  \mid^2}
  &=& \frac{1- \mid q/p \mid^2}{1+ \mid q/p\mid^2}, \mbox{and}  \nonumber \\
\\
tan \phi &=& \frac{Im (q/p)}{Re (qp)} \nonumber
\end{eqnarray}
For the $D^0 - \bar{D}0$ system, the SM values are
\begin{eqnarray}
2Re \ \epsilon_D \approx \frac{1}{2} \ \frac{\delta \Gamma}{\dM}
\left [
\frac {Im \Gamma_{12}}{Re \ \Gamma_{12}} -
\frac{Im \ M_{12}}{Re M_{12}}
\right ]  & \sl & 5.10^{-3}
\end{eqnarray}
and tan$\phi \approx -Im M_{12} /Re M_{12}   \sl  10^{-2}$.
The phase angle $\phi$ is convention dependent and not measurable; but accessible 
in combination with amplitude phases.

There are several ways to measure $Re \epsilon_D$:  (i)comparing the time integrated rates
for $D^0$ and $\bar{D}^0$ to a CP eigenstate final state, the asymmetry
$A = (\Gamma -\bar{\Gamma}) /(\Gamma + \bar{\Gamma}) \cong Re \ \epsilon_D$
(ii) the charge asymmetry in $e^+e^- \ra D^0 \bar{D}^0 \ra \ell^+ \ell^+ x,
\ell^- \ell^- x, \ a = (N^{++} -N^{--})/(N^{++} -N^{--}) \cong \ 4 Re \ \epsilon_D$.

The time dependent decay rates into flavor specific states for states starting as $D^0$
(and $\bar{D}^0)$ are interesting and useful.  The modes into 
$K^+ \pi^-$ and $K^- \pi^+$ have been much discussed recently \cite{Liu}.  
One interesting result is that
if new physics enhances both $\dM$ and $\phi$, then the difference in the rates
$\Gamma (D^0 \ra K^+ \pi^- (t)) - \bar{\Gamma} (\bar{D}^0 \ra K^- \pi^+ (t))$ is proportional
to $(sin \ \phi) \dM t$ at short times and this linear time dependence should be ``easy''
to disentangle.

\section{Rare Decays}

The flavor changing radiative decay which is analogs of the famous 
$b \ra \ s \gamma$ is
$c \ra u \gamma$.  The bare electro-weak penguin for $c \ra u \gamma$ yields a branching
ratio of $10^{-17}$ which is enhanced to $10^{-12}$ by QCD corrections \cite{Burdman2}.  
This would seem
to leave a large window for new physics contributions.  Unfortunately, long distance effects
are very large and close this window \cite{Burdman3}.  
Conventional nearby poles make rates for decays like 
$D \ra \rho \gamma$ in the 
range $10^{-4} - 10^{-6}$.  Hence both the Penguin as well as any new physics are
completely masked by these long distance effects.  Similar long distance effects 
plague \cite{Babu1} decays
with off-shell photons such as $D \ra \ell^+ \ell^- x$.  However, observation and
study of these decay modes would be very useful in understanding long distance
physics.

There are a number of other rare (one-loop) decay modes of D
which do have extremely small rates when evaluated in SM; thus 
providing a potential window for new physics contributions \cite{Burdman1}.

(i)  $D^0 \ra \mu^+ \mu^-$

	At one loop level the decay rate for $D^0 
\ra \mu^+ \mu^-$ is given by

\be
\Gamma (D^0 \ra \mu^+ \mu^-) \ =
\frac{G_F^4 \ m_W^4 \ f_D^2 \ m_\mu^2 \ m_D \mid F \mid^2}
{32 \pi^3} \sqrt{1-4m^2_\mu / m_D^2}
\ee
where
\begin{eqnarray}
\begin{array}{cl}
F = & U_{us} U_{cs}^* \ (x_s + 3/4 \ x _s^2 \ \ell_n 
x _s)
\\
    & U_{ub} U_{cb}^* \ (x _b + 3/4 \ x_b^2 \ell_n 
x_b)
\end{array}
\end{eqnarray}
and $x_i = m_i^2/m_W^2$.  This yields a branching 
fraction of $10^{-19}$.  There are potentially large long 
distance effects; e.g. due to intermediate states such as
$\pi^0, K^0, \bar{K}^0, \eta, \eta')$ or $(\pi \pi, K 
\bar{K})$ etc. Inserting the known rates for $P_i 
\ra \mu^+ \mu^-$ and ignoring the extrapolation the 
result for $B(D^0 \ra \mu^+ \mu^-)$ 
is $3.10^{-15}$.
This is probably an over-estimate but gives some idea 
of the long distance enhancement.

(ii)  $D^0 \ra \gamma \gamma$

The one loop contribution to $D^0 \ra \gamma \gamma$ can be 
calculated in exactly the same way as above and the 
amplitude A is found to be approximately $4.6.10^{-14}$ 
GeV, where $A$ is defined by the matrix element
$A \ q_{1 \mu} \ q_{2\nu} \ \epsilon_{1\rho} \ \epsilon_{2 \sigma} 
\ \epsilon^{\mu \nu \rho \sigma}$.

The decay rate is $\Gamma = \mid A \mid ^2m_D^3 / 64 \pi$ 
and the branching fraction is $10^{-16}.$  The single 
particle contributions due to ($\pi, K, \eta, \eta')$ 
yield $3.10^{-9}$ but again are probably over estimated.

(iii) $D \ra \nu \bar{\nu} x$.

The decay rate for $c \ra u \nu \bar{\nu}$ (for 3 
neutrino flavors) is given by

\begin{equation}
\Gamma = \frac{3 G^2_F \ m_c^5}{192 \pi^3} \left [
\frac{\alpha}{4 \pi x_w} \right ]^2 \mid A_\nu \mid^2.
\end{equation}
Inserting the one loop value for $A_\nu$, one finds for the 
branching fractions:
\begin{eqnarray}
\begin{array}{lcl}
B(D^0 \ra \nu \bar{\nu} x) & = &
2.10^{-15} \\ 
B(D^+ \ra \nu \bar{\nu} x) & = & 4.5.10^{-15}
\end{array}
\end{eqnarray}

For the exclusive modes $D^0 \ra \pi \nu
\bar{\nu}$ and $D^+ \ra \pi^+ \nu \bar{\nu}$ an 
estimate of the long distance contributions yields
\begin{eqnarray}
\begin{array}{lcl}
B(D^0 \ra \pi^0 \nu \bar{\nu}) & \sim  & 5.6.10^{-16}  
\\
B(D^+ \ra \pi^+ \nu \bar{\nu}) & \sim & 
8.10^{-16}
\end{array}
\end{eqnarray}

(iv) D $\ra \bar{K}(K) \nu \bar{\nu}$

These modes have no short distance one loop contributions.  
Estimates of long 
distance contributions due to single particle poles yield 
branching fractions of the order of $10^{-15}.$

\section{Direct CP Violation}
Simplest examples of direct CP violation are rate asymmetries for $D^+$ and $D^-$
decays into charge conjugate final states.  As in now well-documented \cite{Brown}, to obtain
non-zero asymmetries one needs 
i) at least two strong interaction eigenstates in the final
state (e.g. isospins) with unequal final state interaction phases and 
ii) with unequal weak CP phases.  The important and crucial feature of SM is that these
conditions are satisfied only in the Cabibbo-suppressed modes.  Hence no CPV rate asymmetry is
expected for Cabibbo-favored modes (e.g. $D^+ \ra K^- \pi^+ \pi^+)$ or for doubly-Cabibbo-
suppressed modes (e.g. $D^+ \ra K^+ \pi^0)$. 
For the Cabibbo-suppressed modes the asymmetry can be no larger than 
of order $10^{-3}$; $D \ra \ \rho \pi$ seems to be a promising candidate according
to some recent estimates \cite{Buccella}.

\section{New Physics Scenarios}

(i)  Additional Scalar Doublet

One of the simplest extensions of the standard model is to 
add one scalar Higgs doublet \cite{Abbott}.  If one insists on flavor 
conservation there are two possible models:  in one (model 
I) all quarks get masses from one Higgs (say $\phi_2)$ 
and the other $\phi_1$ does not couple to fermions; in the 
other $\phi_2$ gives  
masses to up-quarks only and $\phi_1,$ to down-quarks only.  
The new unknown parameters are $\tan \beta (= v_1/v_2$, the 
ratio of the two vevs) and the masses of the additional 
Higgs scalars, both charged as well as neutral.

In the charmed particle system, the important effects are in 
$\delta m_D$ and the new 
contributions due to charged Higgses to rare decays such as
$D^0 \ra \mu^+ \mu^-, D \ra \pi \ell 
\bar{\ell}, D \ra \gamma \gamma, D \ra \rho 
\gamma$ etc.

The mass of the charged Higgs is constrained to be above 
50 GeV by LEP data and there is a joint constraint on $m_H$ 
and $\tan \beta$ from the observation of $b \ra s \gamma$.  
For large $\tan \beta, \ \delta m_D$ can be larger 
than the SM results \cite{Hewett,Burdman1}.

(ii)  Fourth Generation

If there is a fourth generation of quarks, accompanied by a 
heavy neutrino $(M_{N0} > 50$ GeV to satisfy LEP 
constraints) there are many interesting effects observable in 
the charm system.

In general $U_{ub'}$ and $U_{cb'}$ will not be zero and then 
the $b'$-quark can contribute to $\delta m_D$ as well as to rare 
decays such as $D^0 \ra \mu \bar{\mu}, D \ra 
\ell \bar{\ell} x, D \ra \pi \nu \bar{\nu}$ etc. (A 
singlet b' quark as predicted in E6 GUT has exactly the same 
effect).  A heavy fourth generation neutrino $N^0$ with 
$U_{eN0} U^*_{\mu N} \neq 0$ engenders decays such as $D^0 
\ra \mu \bar{e}$ as well.

For $U_{ub'} U_{cb'} \stackrel{\sim}{>} 0.01$ and $m_{b'}> 
100 GeV$, it is found that \cite{Babu2}
\begin{enumerate}
\item[(a)]$\delta m_D/\Gamma \ \ > 0.01$;
\item[(b)] $B(D^0 \ra \mu \bar{\mu}) > 
0.5.10^{-11}$;
\item[(c)]B $(D^+ \ra \pi^+ \ell \bar{\ell} \ ) >
10^{-10}$; etc. 
\end{enumerate}

For a heavy neutrino of mass $M_{N^0} > 45$ GeV, 
the mixing with $e$ and 
$\mu$ is bounded by $\mid U_{Ne} U^*_{N \mu} \mid^2 < 
7.10^{-6}$  
and we find \cite{Acker} that branching fraction for $D^0 \ra 
\mu^- e^+, \mu^+ e^-$ can be no more 
than $6.10^{-22}$!  This is also true for a singlet heavy 
neutrino unaccompanied by a charged lepton.  To turn this 
result around, any observation of $D^0 \ra \mu e$
at a level greater
than this must be due to some other physics, e.g. a 
horizontal gauge (or Higgs) boson exchange.
\begin{description}
\item[(iii)] Singlet $Q=2/3$ Quarks

In this case, there is a new contribution to $\dM$ at the tree level due to FCNC
coupling to $Z$ giving \cite{Branco}
\be
\delta M_D = \frac{\sqrt{2}}{3} G_F \ f_D^2 \ B_D \ m_D \eta_{QCD} \ \lambda
\ee
where $\lambda = \sum_{i=1}^{3} V_{ui} V^*_{ci}$ indicates the lack of
unitarity of the 3x3 KM matrix.  $\dM_D$ can be as large as $10^{-15} GeV.$
The angle $\phi$ is given by
\be
tan \phi \cong \frac{Im (V_{ub} V_{cb})^2}
{Re (V_{ub} V_{cb}*)^2}
\ee
and can be large.  This form for $tan \phi$ is valid in several scenarios, including
those with charged Higgses.

\item[(iv)] Flavor Changing Neutral Higgs

It has been an old idea that if one enlarges the Higgs 
sector to share some of the large global flavor symmetries 
of the gauge sector (which eventually are broken 
spontaneously) then it is possible that interesting fermion 
mass and mixing pattern can emerge.  It was realized early \cite{Pakvasa1}
that in general this will lead to flavor changing neutral 
current couplings to Higgs.  As was 
stressed \cite{Pakvasa2} then and has 
been emphasized recently \cite{Hall}, this need not be alarming as long 
as current limits are satisfied.  But this means that the 
Glashow-Weinberg criterion will not be satisfied and the GIM 
mechanism will be imperfect for coupling to scalars.  This 
is the price to be paid for a possible "explanation" of 
fermion mass/mixing pattern.  Of course, the current 
empirical constraints from $\delta m_K, K_L \ra
\mu \mu  \ K_L \ra \mu e$ etc. must be observed.  
This is not at all difficult.  For example, in one early 
model, flavor was exactly conserved in the strange sector 
but not in the charm sector!

In such theories, there will be a neutral scalar, $\phi^0$ of 
mass m with coupling such as
\begin{equation}
(g \bar{u} \gamma_5 c \ + \ g' \bar{c} \gamma_5 u) 
\phi^0
\end{equation}
giving rise to a contribution to $\delta m_D$

\begin{equation}
\delta m_D \sim \ \frac{g g'}{m^2} \ f^2_D \ B_D \ m_D 
\left ( m_D/m_C \right )
\end{equation}
With a reasonable range of parameters, it is easily 
conceivable for $\delta m_D$ to be as large as $10^{-13}$ 
GeV.  There will also new contributions to decays such as 
$D^0 \ra \mu \bar{\mu}, D^0 \ra \mu e$ which 
will depend on other parameters.

There are other theoretical structures which are 
effectively identical to this, e.g. composite technicolor.  
The scheme discussed by Carone and Hamilton \cite{Carone} leads to a 
$\delta m_D$ of $4.10^{-15}$ GeV.

\item[(v)] Family Symmetry

The Family symmetry mentioned above can be gauged as well as 
global.  In fact, the global symmetry can be a remnant of an 
underlying gauged symmetry.  A gauged family symmetry leads 
to a number of interesting effects in the charm 
sector \cite{Volkov}.

Consider a toy model with only two families and a 
$SU(2)_H$ family gauge symmetry acting on LH doublets; with
\begin{eqnarray*}
\left [ \left ( \begin{array}{c}
   u \\ c \end{array} \right )  
\left ( \begin{array}{c} d \\ s  \end{array} \right) \right ]_L   \mbox{and}
\left [  \left ( \begin{array}{c}
 \nu_e \\ \nu_\mu \end{array} \right )     
\left ( \begin{array}{c} e \\ \mu \end{array} \right ) \right ]_L \nonumber \\
\end{eqnarray*}
assigned to $I_H = 1/2$ doublets.  The gauge interaction 
will be of the form:

\begin{eqnarray}
g \left [ \overline{(d \ s)}_L \ \gamma_\mu \ \bar{\tau}.
\bar{G} \mu \left (
\begin{array}{c}
d \\ s \end{array} \right )_L \ \ + .......\right ]
\end{eqnarray}
After converting to the mass eigenstate basis for quarks, 
leptons as well as the new gauge bosons, we can calculate 
contributions to $\delta m_K, \delta m_D$ as well as to 
decays such as $K_L \ra e \mu$ and $ D \ra 
e\mu$.  The results depend on
$\theta_d, \theta_u$ and $\theta_e$ which are the unknown mixing 
angles in the $d_L-s_L$, $u_L-c_L$ and $e_L-\mu_L$ sectors 
and the gauge boson
masses.  It is possible to obtain $\delta m_D \sim 10^{-13}$ 
GeV and $B (D^0 \ra e \mu) \sim 10^{-13}$ while 
satisfying the bounds on $\delta m_K$ and $B(K_L^0 
\ra e \mu)$.

\item[(vi)] Supersymmetry

In the Minimal Supersymmetric Standard Model new 
contributions to $\delta m_D$ come from gluino exchange box 
diagram and depend on squark mixings and mass splittings.  
To keep $\delta m_K^{SUSY}$ small the traditional ansatz has 
been squark degeneracy.  In this case $\delta m_D^{SUSY}$ is 
also automatically suppressed, no more than $10^{-18}$ GeV \cite{Hagelin}.  
It has been proposed that 
another possible way to keep $\delta m_K^{SUSY}$ small is to 
assume not squark degeneracy but proportionality of the squark 
mass matrix to the quark mass matrix \cite{Nir}. It turns out in 
this case that $\delta m_D$ can be as large as the current 
experimental limit.  In a very recent proposal of ``effective
supersymmetry'' which is a new approach to the problem of FCNC in supersymmetry,
there could be also significant contributions to $\delta m_D$ \cite{Cohen}.

\item[(vii)]  Left-Right Symmetric Models

In general Left-Right symmetric theories do not lead to interesting
predictions for the D system.  There is one exception:  as pointed out by the Orsay
group \cite{Yaouanc}, it is possible to obtain sizable direct 
CPV rate asymmetries in Cabibbo-allowed
modes.
\end{description}

\section*{Conclusion}

My personal prejudice is that if we are to understand
fermion mass/mixing pattern at accessible energy scales, then GIM violation and FCNC must exist; 
and the charm system offers the largest window of opportunity for this search.
They must be found!

\section*{Acknowledgment}

Most of the work reported here is based on on-going collaboration with Tom Browder, Gustavo
Burdman, Eugene Golowich and JoAnne Hewett, and is supported in part by U.S.D.O.E under 
Grant DE-FG 03-94ER40833.

\end{document}